%Paper: hep-th/9411140
%From: Tony Giaquinto <Tony.Giaquinto@math.lsa.umich.edu>
%Date: Thu, 17 Nov 1994 18:42:14 -0500

\input amstex
\magnification 1200
\documentstyle{amsppt}
\NoRunningHeads
\pageheight {7.291in}
\pagewidth {5.5in}
\define \ts{\otimes}
\define \ps{k[[t]]}
\define \ra{\rightarrow}
\define \U{U\frak g}
\redefine \D{\Delta}
\redefine \d{deformation }

\redefine \I{\text {Id}}
\redefine \d{\cdot}
\redefine \l{\langle}
\redefine \r{\rangle}
\TagsOnRight
\title  BIALGEBRA ACTIONS, TWISTS, AND UNIVERSAL DEFORMATION FORMULAS
     \endtitle
\author Anthony Giaquinto and James J. Zhang
\endauthor
\address  Department of Mathematics, University of Michigan, Ann
Arbor MI 48109-1003
    \endaddress
\curraddr (J.J.Z.) Department of Mathematics, University of Washington,
Seattle, WA 98195
\endcurraddr
\email tonyg\@math.lsa.umich.edu (A.G.) and
     jzhang\@math.washington.edu (J.J.Z.)
     \endemail
\thanks The authors wish to thank the NSA and NSF
for partial support of this work. Both authors thank
M. Gerstenhaber, S.D. Schack, J. Stasheff and J.T. Stafford for useful
conversations about this work.\endthanks

\keywords  twist, Hopf-algebras, universal deformation formulas
     \endkeywords

\abstract  We introduce the concept of a twisting element based
on a bialgebra and show how it can be used to twist a large class of algebras,
coalgebras and certain subcategories of their respective module
and comodule categories. We prove that this subcategory of modules
over the original algebra is equivalent to the corresponding category
of modules over the twisted algebra. The relation between twisting
elements and universal deformation formulas is also given, along with
new formulas which are associated to enveloping algebras of non-abelian
Lie algebras.

\endabstract
\endtopmatter

\document
\baselineskip 16pt
\head {\bf 0. Introduction} \endhead
The purpose of this paper is to introduce a general
notion of ``twisting'' algebraic structures
based on actions of a bialgebra $B$ and to relate these
twists to deformation theory. We show that
elements of $B\ts B$ can be used to twist the multiplication of any
$B$-module algebra or $B$-module coalgebra. Moreover,
we focus on a certain subcategory
of left $A$-modules which naturally twists to a subcategory of modules
over the twisted algebra and we show that these two categories are
equivalent. Similar statements can be made for certain categories
of comodules. When the bialgebra is
$(kG)^*$ with $G$ a finite group or monoid, the algebras to which our
construction applies are the
$G$-graded algebras $A=\bigoplus _{g\in G}A_g$ (these being the
$(kG)^*$-module algebras) and the
modules which twist are the graded modules.
For algebras graded by an infinite group or monoid it is more
convenient to use the ``dual'' of our theory which will produce twists of
$A$ as a $kG$-comodule algebra. In these cases, the twists we
obtain are precisely the ``cocycle'' twists of [AST].

Another important case is when the bialgebra is the universal enveloping
algebra of a Lie algebra. Here, the twists which naturally
arise are related to deformation theory and, in particular,
``universal deformation formulas'' (cf [GGS, Section 6] or [CGS]).
Our theory of twisting provides alternative proofs to the
basic properties of these formulas. In particular,
we bypass
the cohomological arguments previously used to establish that the deformations
obtained from these formulas are in fact associative.
The supply of such formulas is scarce.
 Indeed, apart from those based on
commutative,
bialgebras only a few isolated examples are known, even though
some results of Drinfel'd ([D]) imply that there is such a formula
associated to every (constant) unitary solution to the classical
Yang-Baxter equation. We broaden the supply of universal deformation
formulas here by providing the first family based on
non-commutative bialgebras.
More specifically,
for every $n\geq 3$ we present a formula based on
the universal enveloping algebra $U\Cal H$ where
$\Cal H \subset {\frak {sl}}(n)$ is a central extension of a
Heisenberg Lie algebra. These formulas
are generalizations of the ``quasi-exponential'' formula based on
the two-dimensional solvable Lie algebra, see [CGG].
A natural use of a universal deformation formula
based on an enveloping algebra $\U$
is to deform coordinate rings of various algebraic varieties. In particular,
if an algebraic group with Lie algebra $\frak g$ acts as automorphisms
of a variety $V$, then the formula induces a deformation of
the coordinate ring $\Cal O(V)$. That is, the quantization takes
place solely because of the action of the group.
These quantizations are generalizations of the
classical ``Moyal product'' ([M]) which, in our language
is a deformation of $\Cal O(\Bbb R^{2n})$
obtained from a universal deformation
formula associated to an abelian Lie algebra. This quantization is
in the ``direction''
of the standard Poisson bracket on $\Cal O(\Bbb R^{2n})$.
In a similar way, a universal deformation formula
provides deformations of coordinate rings in the direction of
a suitable Poisson bracket. In the analytic
case, if $G$ is a Lie group which
acts as diffeomorphisms of a manifold
$M$, then a formula based on $U\frak g$ will deform
the ring $C^{\infty}(M)$. For actions of $\Bbb R^d$, Rieffel has shown
more --  namely that this action induces a strict deformation quantization
of $C^{\infty}(M)$ as a $C^*$-algebra. It is natural to ask whether
similar constructions exist for actions of non-abelian Lie groups.

Finally, we give here a skew form of the
``quasi-exponential'' formula mentioned above.
A skew form of a universal deformation
formula has certain parity
properties that make it more convenient in the study of $*$-products on
Poisson manifolds. We do not, as of yet, have a skew form for
our formula based on $U\Cal H$ although, in theory, every formula based on
an enveloping algebra can be skew-symmetrized.
\head {\bf 1. Twisting and Deformations} \endhead
Let $k$ be a fixed commutative ring. Throughout this paper, all
algebras, coalgebras, and their respective modules and comodules will
be symmetric $k$-modules and their tensor product over $k$ will
be denoted $\ts$. We will generally follow [Mon] for basic definitions
and notation about bialgebras and their actions.
Suppose that $B=B(\mu_B,\D_B,1_B,\varepsilon_B)$ is a $k$-bialgebra
with multiplication $\mu_B:B\ts B\ra B$, comultiplication
$\D_B:B\ra B\ts B$, unit $1_B$, and counit
$\varepsilon _B: B\ra k$. If $M$ is a left $B$-module and $b\in B$, then the
left multiplication map sending $m\in M$ to $b\cdot m$
will be denoted
$b_l :M\ra M$. For a right $B$ module $N$, we similarly have the right
multiplication map $b_r:N\ra N$. In this situation, we can make
$M\ts M$ and $N\ts N$ into, respectively, left and right $B\ts B$-modules
via the maps
$(b\ts b')_l :M\ts M\ra M\ts M$ and $(b\ts b')_r:N\ts N\ra N\ts N$
which
send $m\ts m'$ to $(b\d a)\ts (b'\d a')$ and
$n\ts n'$ to $(n\d b)\ts (n'\d b')$.
\definition{Definition 1.1} A left $B$-module $A$ is a {\it {left $B$-module
algebra}} if $A=A(\mu_A,1_A)$
is a unital $k$-algebra such that for all $b\in B$,
\roster
\item $b\d 1_A=(\varepsilon (b))\d 1_A$\qquad and
\item $b\d (aa')=\sum (b_{(1)}\d a)(b_{(2)}\d a')$ where
$\D_B(b)=\sum b_{(1)}\ts b_{(2)}$.\endroster \enddefinition
Condition (1.1.2) is equivalent to commutativity of
the diagram
$$\CD
A\ts A@> \mu_A > > A\\
@V(\D_B(b))_l VV  @Vb_l VV\\
A\ts A@> \mu_A > > A \endCD$$
for all $b\in B$. It is also easy to see that
if $A$ is a left $B$-module algebra then the primitive elements of
$B$ act as derivations of $A$ and the grouplike elements act
as automorphisms.
In a similar way, we can define the notion
of a right $B$-module algebra.
The foregoing may be easily dualized
to obtain the notion of a left or right $B$-module coalgebra.
\definition{Definition 1.2} An element $F\in B\ts B$ is a
{\it {twisting element}} (based on $B$) if
\roster
\item
$(\varepsilon_B \ts \I)F=1\ts 1 =(\I \ts \varepsilon_B)F$, \quad and
\item $\left[(\D_B\ts \I )(F)\right](F\ts 1)=
\left[(\I \ts \D_B)(F)\right](1\ts F).$ \endroster
\enddefinition
Note that (1.2.2) is an expression which must hold
in $B\ts B\ts B$.
As we shall see, such $F$ can be used to ``twist'' the multiplication
of any left $B$-module algebra $A$ and the comultiplication of any
right $B$-module coalgebra $C$. More precisely, a twisting element provides
a new multiplication on the underlying $k$-module of a
$B$-module algebra and a new comultiplication on the underlying
$k$-module of a $B$-module coalgebra.
For the algebra case, the twisted multiplication
is defined to be the composite $\mu_ A\circ F_l :A\ts A\ra A$
and for the coalgebra case the twisted comultiplication is
$ F_r\circ \D_C:C\ra C\ts C$.
\proclaim{Theorem 1.3} Let $F\in B\ts B$ be a twisting element.
\roster
\item If $A$ is a left $B$-module algebra, then $A_F=A(\mu_A\circ F_l,1_A)$
is an associative $k$-algebra.
\item If $C$ is a right $B$-module coalgebra, then
$C_F=C(F_r\circ \D_C, \varepsilon_C)$
is a coassociative $k$-coalgebra.
\endroster \endproclaim
\demo{Proof}
We only prove (1) since its dual establishes (2).
The associativity of $\mu_A \circ F_l$ can be determined by considering
the following diagram:
$$\CD
A\ts A\ts A @>  F_l \ts \I> > A\ts A\ts A@>  \mu_A\ts \I> > A\ts A\\
@V \I\ts F_l  VV @V ((\D_B\ts \I)F)_l VV @V F_l VV\\
A\ts A\ts A @>  ((\I \ts \D_B)F)_l > > A\ts A\ts A@>  \mu_A\ts \I> > A\ts A\\
@V \I\ts \mu_A VV @V \I \ts \mu_A VV @V \mu_A VV\\
A\ts A@>  F_l> > A\ts A @>  \mu_A> > A. \endCD$$
Each of the four inner squares commutes:
the lower right square represents the associativity of $\mu_A$, the top left
square commutes by (1.2.2) since we are assuming
that $F$ is a twisting element, and the commutativity of the
two ``off-diagonal'' squares follows from (1.1.2)
since $A$ is a left $B$-module algebra.
Consequently, the outer square commutes. Now the composite of
the far right column with the top row
is the map $(\mu_A \circ F_l)((\mu_A \circ F_l)\ts \I)$
while the composite of the bottom row with the
far left column is
$(\mu_A \circ F_l)(\I\ts (\mu_A \circ F_l))$ and thus
$\mu_A \circ F_l :A\ts A\ra A$ is associative. To see that $1_A$ remains
the unit under this new multiplication, note that by (1.1.1) and
(1.2.1) we have
$[\mu_A \circ F](1_A\ts a)=[\mu_A\circ (\varepsilon_B\ts \I) F] (1_A\ts a)
=\mu_A(1_A\ts a)=a$.
Similarly, $1_A$ also serves
as the
right unit under $\mu_A \circ F$.
$\blacksquare$
\enddemo
\remark{Remarks 1.4}
\roster
\item In the same manner,
if $P\in B\ts B$ satisfies
$$(P\ts 1)\left[(\D_B\ts \I)(P)\right]=(1\ts P)\left[ (\I\ts \D_B)(P)\right]$$
then $P$ will twist any right $B$-module
algebra and any left $B$-module coalgebra.
\item If $A=\bigoplus _{g\in G}A_G$ is a graded algebra with $G$ a finite
group or monoid,
(i.e. $A$ is a left $(kG)^*$-module algebra) and
$F$ is a twisting element based on $(kG)^*$, then $A_F$
is a cocycle twist in the sense of [AST]. When $G$ is infinite,
we instead can use a dual construction and twist $A$ as a $kG$-comodule
algebra.
The twists of graded algebras appearing
in [Z] are, in general, not obtainable by twisting elements.
\endroster
\endremark
A natural question is whether a twisting element can be used to twist
bialgebras. So far, we are unaware of any general method of twisting
both the multipliction and comultiplication of a bialgebra. There is,
however, a notable exception: it is possible twist the
comultiplication of $B$ itself in such a way
that it remains compatible with its original
multiplication, unit, and counit. Moreover, this twisted bialgebra
acts in a natural on the twisted algebras $A_F$ and coalgebras $C_F$
formed using Theorem 1.3.
To establish this we use the fact that
the right multiplication map $b_r:B\ra B$ sending $x$ to $xb$ gives $B$
the structure of a right $B$-module coalgebra. Similarly, $B$ is also
a left $B$-module coalgebra via left multiplication $b_l:B\ra B$.
\proclaim{Theorem 1.5} Let
$F$ be an invertible twisting element based on a bialgebra $B$.
\roster
\item  If $\D'_B=F^{-1}_l\circ F_r \circ \D_B$, then
$B_F=B(\mu_B,\D_B',1_B,\varepsilon_B)$ is a $k$-bialgebra.
\item If $A$ is a left $B$-module algebra then $A_F$ is a left
$B_F$-module algebra.
\item If $C$ is a right $B$-module coalgebra then $C_F$ is a right
$B_F$-module coalgebra. \endroster
\endproclaim
\demo{Proof} By Theorem 1.3.2, we have that $ F_r \circ \D_B:B\ts B\ra B$ is
coassociative since $F$ is a twisting element.
This comultiplication, however, is generally not compatible with $\mu_B$, but
it
further twists to one which is.
To see this, first note that inverting both sides of
$$\left[(\D_B\ts \I)F\right](F\ts 1)=\left[(\I\ts \D_B)F\right](1\ts F)$$
yields
$$(F^{-1}\ts 1)\left[(\D_B \ts \I)F^{-1}\right]=
(1\ts F^{-1})\left[(\I\ts \D_B)F^{-1}\right].\tag 1.6$$
Now the twisted coalgebra $B$ (with comultiplication $ F_r \circ \D_B$)
remains a left $B$-module coalgebra since $(F_r \circ \D_B) (b)=
(\D_B(b))F$ only involves right multiplication by $F$.
In light of (1.6) and Remark 1.4, it follows that
$\D_B'=F^{-1}_l \circ F_r \circ \D_B$ is coassociative. For $b\in B$,
we have that $\D'_B(b)$ is just the conjugate $F^{-1}\D_B(b)F$
and so this comultiplication is compatible with $\mu_B$.
Moreover, since the algebra structure remains unchanged,
the original counit $\varepsilon_B:B\ra k$ is
an algebra map and consequently, $B(\mu_B,\D_B',1_B,\varepsilon_B)$ is a
bialgebra.\newline
To prove (1.5.2), we first need to establish that
$A_F$ is actually a left $B_F$ module.
To do so, we define the action of $b\in B_F$ on $a\in A_F$ to be just
$b\cdot a$, the action given by the original $B$-module structure of $A$.
This action is well-defined since $A$ and $B$ coincide as $k$-modules with
$A_F$ and $B_F$ and the multiplications of $B$ and $B_F$ are identical.
With this module structure, (1.1.1) is automatically satisfied since
the counits of $B$ and $B_F$ are also identical.
To show that condition (1.1.2) holds we need to use the twisted
multiplication in $A_F$ and twisted comultiplication in $B_F$.
Now the map $(\mu_A\circ F_l) \circ (F^{-1}\D_B(b)F)_l :A\ts A\ra A$
is the same as $(\mu_A \circ (\D_B(b))_l) \circ F_l$. The latter map
can be expressed as $(b_l\circ \mu_A)\circ F_l$ as $A$ is a left $B$-module
algebra. Hence we have the equality
$(\mu_A\circ F_l) \circ (F^{-1}\D_B(b)F)_l=b_l\circ (\mu_A\circ F_l)$
and so $A_F$ is a left $B_F$-module algebra.
\newline The (dual) proof of (1.5.3) is omitted.
$\blacksquare$
\enddemo
If $B$ is commutative then the twist $B_F$ is just $B$ itself and so
in this case $A_F$ and $C_F$ are still, respectively, a left $B$-module
algebra and right $C$-module coalgebra.

We now turn to the topic of twisting modules and comodules along with
algebras and coalgebras. Only the details for the module/algebra case
are given here as those for the comodule/coalgebra case are easily
obtainable by dualization. If $M$ is a left $A$-module and $A_F$ is
a twist of $A$, it is natural to ask whether $M$ twists to a left
$A_F$-module. This is, in general, not possible as we must take into
account the action of the bialgebra $B$. We will
be concerned with the following class of modules.
\definition{Definition 1.7}Suppose that $A$ is a left $B$-module and
let $A$-{\bf{Mod}} be the category of all left $A$-modules. Define
$(B,A)$-{\bf{Mod}} to be the subcategory of
$A$-{\bf{Mod}} whose objects are those
left $A$-modules $M$ which are also left $B$-modules
which satisfy $b\cdot (a\cdot m)=\sum (b_{(1)}\cdot m)(b_{(2)}\cdot a)$ for
all $b\in B$, $a\in A$, and $m\in M$. Morphisms in $(B,A)$-{\bf{Mod}}
are $k$-linear maps which are simultaneously $A$-module and $B$-module
maps.\enddefinition
If $A=\bigoplus _{g\in G}A_g$ is a $G$-graded algebra then
$((kG)^*,A)$-{\bf{Mod}} is the
familiar category
of graded left $A$-modules. These are left $A$-modules $M$ with
$M=\bigoplus_{g\in G}M_g$ as a $k$-module and $a_g\cdot m_h \in M_{gh}$ for all
$a_g\in A_G$ and $m_h\in M_h$.

Now if $M\in (B,A)$-{\bf{Mod}}, then it is possible to define
a twist $M_F$ of $M$ which will be an $A_F$-module. Just as
$A_F$ and $A$ coincide as $k$-modules, the modules $M_F$ and $M$ are
also identical $k$-modules.
To define this twist
we use the fact that $M\ts A$ is a left $B\ts B$-module since $A$ and $M$
are both left $B$-modules. In particular, for
$F=(\sum f_i\ts f'_i)\in B\ts B$, the map $F_l:A\ts M\ra A\ts M$ sends
$(a\ts m)$ to $\sum (f_i\cdot a) \ts (f'_i\cdot m)$. If
$\lambda :A\ts M\ra M$ is the original left $A$-module structure map then
the twisted module structure map is $\lambda_F=\lambda \circ
F_l:A_F\ts M_F \ra M_F$. It is not hard to see that $M_F$ is a well-defined
$A_F$-module. Indeed,
the equality
$\lambda _F(a\ts \lambda_F(a'\ts M))=
\lambda_F( \mu_A\circ F_l(a\ts a')\ts m)$ follows from (1.2.2) and the
fact that $b\cdot (a\cdot m)=\sum (b_{(1)}\cdot m)(b_{(2)}\cdot a)$ while
$\lambda _F(1_A\ts m)=m$ follows from (1.1.1) and (1.2.1).
Now since the bialgebra $B_F$ has the same algebra structure
as $B$, the twisted module $M_F$ is also a left $B_F$-module under
the original action of $B$ on $M$. Thus $M_F$ is both a left $A_F$-module
and a left $B_F$-module and  it is not hard to
verify that these actions are compatible in the
sense that $M_F\in (A_F,M_F)$-{\bf {Mod}}.
\proclaim{Theorem 1.8} If $F\in B\ts B$ is an invertible twisting element
and
$A$ is a left $B$-module algebra, then
the categories
$(B,A)$-{\bf{Mod}} and $(B_F,A_F)$-{\bf{Mod}} are equivalent.
\endproclaim
\demo{Proof}
The rule assigning each
$M$ to the twist $M_F$
defines a functor
$\Cal F:(B,A)\text{-\bf{Mod}} \longrightarrow (B_F,A_F)\text{-\bf{Mod}}$.
This functor has a two-sided inverse because we can twist
$A_F$ and $B_F$ back to their untwisted versions. The reason
why this is possible is that $F^{-1}$ is a twisting element
based on $B_F$. To establish this we need to show that if
$F^{-1}=\sum g_i\ts h_i$ then
$$\left[\sum (F^{-1}\D_B(g_i)F)\ts h_i\right](F^{-1}\ts 1)=
\left[\sum g_i\ts (F^{-1}\D_B(h_i)F)\right](1\ts F^{-1}).$$
This equation is equivalent to having
$$(F^{-1}\ts 1)[(\D_B\ts 1)F^{-1}]=(1\ts F^{-1})[(1\ts \D_B)F^{-1}].$$
Now the latter equation is satisfied since inverting both sides
yields (1.2.2) which holds since $F$ is a twisting element based on $B$.
We may thus form  $(A_F)_{F^{-1}}$ and $(B_F)_{F^{-1}}$ which
are clearly just the original $A$ and $B$ and, in the same way as above,
we get a functor
$ \Cal F^{-1}:(B_F,A_F)\text{-\bf{Mod}} \longrightarrow (B,A)\text{-\bf{Mod}}$
which is clearly a two-sided inverse to $\Cal F$ and so
these two categories are equivalent. $\blacksquare$ \enddemo

There are remarkably few techniques to explicitly produce twisting
elements. In some cases though, it is possible to multiply
two twisting elements to produce a new one.
\proclaim{Proposition 1.9} Suppose that $B'$ and $B''$ are commuting
sub-bialgebras of
$B$ and let $F\in B'\ts B'$ and $F'\in B''\ts B''$
be twisting elements based on
$B'$ and $B''$, respectively. Then
$FF'$ is a twisting element based on the smallest sub-bialgebra
of $B$ which contains $B'$ and $B''$.\endproclaim
\demo{Proof} Since elements of $B'$ commute with those of
$B''$ we have that
$$[(\D \ts 1)(FF')](FF'\ts 1)=[(\D \ts 1)F](F\ts 1)[(\D \ts 1)F'](F'\ts 1).$$
Because $F$ and $F'$ are UDF's, this may be expressed as
$$[(1 \ts \D)F](1\ts F)[(1 \ts \D)F'](1\ts F')$$
which in turn equals
$$[(1\ts \D)(FF')](1\ts FF').\, \,  \blacksquare $$
\enddemo
For any bialgebra $B$, the element $1_B\ts 1_B$ is obvioulsy a twisting
element; when used as in Theorems 1.3 or 1.5,
it is the ``identity twist'', that is, it effects no change.
Apart from this
trivial example and the cocycle twists previously mentioned,
the class of quasi-triangular bialgebras gives other twisting elements.
\definition{Definition 1.10 [D2]} A bialgebra $B$ is {\it quasi-triangular} if
there is an invertible element $R=\sum a_i\ts b_i \in B\ts B$ such that
\roster
\item $\D_B(b)=R^{(-1)}(\D_B^{ {\text {op}}}(b))R$ for all $b\in B$ where
$\D_B^{ {\text {op}}}(b)=\sum b_{(2)}\ts b_{(1)}$,
\item $(\D_B\ts \I)(R)=R_{13}R_{23}$,\quad and
\item $(\I\ts \D_B)(R)=R_{13}R_{12}$
\endroster
where $R_{12}=R\ts 1,$ $R_{23}=1\ts R$, and $R_{13}=\sum_i a_i\ts 1\ts b_i$.
\enddefinition
A basic fact about quasi-triangular bialgebras is that the associated
$R$ satisfies the quantum Yang-Baxter equation
$$R_{12}R_{13}R_{23}=R_{23}R_{13}R_{12}.\tag 1.11$$
This fact, together with
(1.10.2), (1.10.3), and Remark 1.4 imply that $R^{-1}$ is a twisting element.
\example{Example 1.12}
\roster
\item (Majid [Ma])
Let $G=\Bbb Z/(n\Bbb Z)$ be the cyclic group of order $n$ with generator $g$.
If $n$ is invertible in $k$ and $q\in k$ is a primitive nth root
of unity then then the group
bialgebra $K[G]$ then has a non-trivial quasi-triangular structure in which
$$R=\frac {1}{n} \sum_ {i,j=0}^{n-1} (-1)^{i+j}q^{i+j}g^i\ts g^j.$$
\item (Drinfel'd [D2])
If $B$ is a finite dimensional bialgebra then its ``double''
$D(B)$ is a quasi-triangular bialgebra. As a $k$-module, $D(B)=B^*\ts B$;
its bialgebra structure is intricate and not necessary for this
paper and we refer the reader to [D2] or [Mon] for a careful
discussion of this topic.
\item The classic example of a quasi-triangular bialgebra is the
quantized enveloping algebra $U_q\frak g$, see [D2]. To obtain
its quasi-triangularity, it is necessary to consider $U_q\frak g$
as a $\ps$-module with $q=e^t$ and take its completion with respect to the
$t$-adic topology. The corresponding $R$ is then expressible as an infinite
series in $t$.
\endroster
\endexample

Our main interest in the remainder of this paper will be
twisting elements which can be used to obtain deformations
of various algebraic structures and so we will consider
those based on $k[[t]]$-bialgebras.
We now briefly recall the main
aspect of formal algebraic deformations.
\definition {Definition 1.13} Let $A$ be an
associative $k$-algebra with multiplication
$\mu _A:A\ts A\ra A$. Then a
{\it {formal deformation}} of
$A$ is a $k[[t]]$-algebra structure on the power series
module $A[[t]]$ with multiplication
of the form
$$\mu _A' =\mu _A+t\mu _1+t^2\mu _2+\cdots + t^n\mu _n+\cdots$$
where each
$\mu _i:A \times A\rightarrow A$ is a $k$-bilinear map
extended to be $k[[t]]$-bilinear.
\enddefinition
\remark{Remarks 1.14}
\roster
\item Note that $\mu _A'$ is completely determined by its effect on
$A\ts A$. Even though this a formal construction it turns out that
in many cases
$\mu_A'(a\ts b)$ is actually a polynomial in $t$ and so, by specialization,
we can view the deformation as a $k$-algebra. We can obtain the same
conclusion in many cases by convergence
considerations (when the base field is $\Bbb R$ or $\Bbb C$.)
\item In a more general context it is possible to consider
deformations over other rings in the
following sense:
If $R$ is a commutative ring with epimorphism $R\ra k$
then a deformation of $A$ over $R$ is
an $R$-algebra $A'$ which is a flat $R$-module together with an isomorphism
$A'\otimes _{R}k\cong A$.
The case $R=k[q,q^{-1}]$ is of particular interest
in the study of quantum groups.
\endroster \endremark
Returning now to formal deformations, we say that deformations $A'$
and $A''$ are equivalent if there is a $\ps$-algebra isomorphism
$\phi :A'\rightarrow A''$ of the form $\phi = {\text {Id}}_A +
t\phi _1 +t^2\phi_2 +\cdots $ where $\phi _i :A\rightarrow A$
is a $k$-linear map extended to be $\ps$-linear.
If we set $\mu_0=\mu_A$, then the associativity condition
$\mu _t (\mu _t(a, b), c)=\mu _t(a,\mu _t(b,c))$ is equivalent to having
$$\sum  \Sb i+j=n \\ i,j\geq 0 \endSb \mu _i(\mu _j(a,b),c)-
\mu _i(a,\mu _j(b,c))=0\tag 1.15$$
for all $n\geq 0$ and $a,b,c \in A$. In particular,
when $n=1$ we have that the {\it {infinitesimal}},
$\mu _1$, must satisfy
$$\mu _1(a,b)c-a\mu _1(b,c)+\mu _1(ab,c)-\mu _1(a,bc)=0$$
and so $\mu _1\in Z^2(A,A)$, the $k$-module of Hochschild 2-cocycles
for the algebra $A$ with coefficients in itself. Equivalent
deformations have cohomologous infinitesimals and so $H^2(A,A)$
may be interpreted as the space of equivalence classes of infinitesimal
deformations of $A$. Given an element of $Z^2(A,A)$, it is natural to ask
whether there is a deformation of $A$ with that infinitesimal.
In general, this is not possible since there may be obstructions,
all lying in $H^3(A,A)$, to ``integrating'' an infinitesimal to
a full deformation. Namely, if $\mu +t\mu _1+t^2\mu _2+\cdots +t^{n-1}\mu
_{n-1}$
defines an associative multiplication on $A[t]/t^n$ then
$$\sum \Sb i+j=n \\ i,j > 0\endSb
 \mu  _i(\mu _j(a,b),c)-\mu _i(a,\mu _j(b,c))$$
is automatically a Hochschild 3-cocycle and must be a coboundary
if the multiplication is extendible to an associative
product on $A[t]/t^{n+1}$.
Even when an infinitesimal $\mu  _1$ is known to be integrable to a full
deformation, e.g. whenever $H^3(A,A)=0$, very little is known
on how to find $\mu  _2, \mu _3, \ldots $ such that
$\mu _t =\sum _{i\geq 0} t^i\mu _i$ satisfies (1.15).

It is straightforward to dualize the deformation theory of algebras
to that of coalgebras. If $C$ is a coalgebra with comultiplication
$\D _C :C\ra C\ts C$, then a formal deformation of $C$ is a
$\ps$-coalgebra structure on $C[[t]]$ with comultiplication of
the form $\D _C '=\D _C+t\D_1+\cdots +t^n\D _n +\cdots $
where each $\D _i:C\ra C\ts C$ is a $k$-linear
map extended to be $k[[t]]$-linear. Note that $\D_C' (c)$ does not generally
lie in $C[[t]]\ts _{k[[t]]} C[[t]]$ and so we must consider its completion,
$(C\ts C)[[t]]$ with respect to the $t$-adic topology. Now
the coassociativity condition
$$(\D_C' \ts 1)\D_C'  =(1\ts \D_C')\D_C'  $$
imposes restrictions on the maps $\D_i$ which may be
interpreted in terms of the coalgebra cohomology of $C$. As in the
algebra case, it is generally difficult to explicitly produce deformations of
a coalgebra.

One way to produce deformations for a wide class
of algebras and coalgebras is by use of certain twisting elements.
\definition {Definition 1.16} A {\it {universal deformation formula}}
(UDF) based on a bialgebra $B$ is a twisting element $F$ based on
$B[[t]]$
of the form
$$F=1\ts 1+tF_1 +t^2F_2 + \cdots +t^nF_n +\cdots$$
where each $F_i\in B\ts B$.\enddefinition
If $A$ is a left $B$-module algebra and $C$ is a right $B$-module coalgebra
then $A[[t]]$ and $C[[t]]$ naturally become a left $B[[t]]$-module algebra
and right $B[[t]]$-module coalgebra, respectively. Now if a UDF
$F$ is based on $B[[t]]$ then the twists of $A[[t]]$ and $C[[t]]$
obtained from Theorem 1.3 are clearly deformations of $A$ and $C$.
Now since any UDF is invertible, it may also be used to twist
the bialgebra $B[[t]]$ according to Theorem 1.5. Since the algebra
structure of $B[[t]]$ is strictly unchanged, this will
be a ``preferred deformation'' of
$B$ in the sense of [GGS].

A fundamental question is to determine ``all'' UDF's based on a bialgebra
$B$.
This must be understood under a natural notion of equivalence. Namely,
if $F$ is a UDF and
$f=1+tf_1+\cdots +t^nf_n +\cdots \in B [[t]]$ then
$\overline{F} =(\D_B f)(F)(f^{-1}\ts f^{-1})$ is also a UDF and we say
$F$ and $\overline{F}$ are equivalent. When used as in Theorem 1.3 or
Theorem 1.5, a UDF produces equivalent deformations. Little is known about the
equivalence
classes of UDF's except when $B=\U$. This case is of greatest interest to us
in part due to its extensive connections to the theory of
quantum groups. For simplicity, we denote the comultiplication of
$\U$ simply
by $\D$ instead of $\D_{\U}$.

Up to equivalence, the existence problem for UDF's based on $\U [[t]]$ has been
completely settled by in [D1], (see [MV] for a more detailed discussion
of the ideas in [D1]).
Drinfel'd also gives an important connection between UDF's and
solutions of the Yang-Baxter equations.
Recall that
$r\in \frak g\ts \frak g$ is a solution to the classical
Yang-Baxter equation if
$$[r_{12},r_{13}]+[r_{12},r_{23}]+[r_{13},r_{23}]=0$$
and $R\in \U \ts \U$ is a solution to the quantum
Yang-Baxter equation if it satifies
(1.11).
Solutions to these equations are {\it {unitary}} if
$r_{12}+r_{21}=0$ in the classical case (i.e. $r$ is skew-symmetric) and
$R_{12}R_{21}=1$ in the quantum case.
\proclaim{Theorem 1.17 (Drinfel'd)} Suppose that
$F=1\ts 1+\sum _{i=2}^\infty t^i F_i$ is a UDF and let
$r=F_1 -(F_1)_{21}$, the skew-symmetrization of $F_1$.
\roster
\item $r$ is a unitary solution to the classical Yang-Baxter equation.
\item If $\overline{F}$ is a UDF equivalent to $F$ then
$F_1 -(F_1)_{21}=\overline{ F}_1 -(\overline {F}_1)_{21}$
\item $F$ is equivalent to a UDF of the form
$1\ts 1 +\frac {1}{2} tr+\sum_{i\geq 2}t^iF_i $ where $F_i=(-1)^i(F_i)_{21}.$
\item $F_{21}^ {-1}\, F$ is a unitary solution to the quantum Yang-Baxter
equation.
\item If $S$ is any unitary solution to the classical
Yang-Baxter equation then there is a UDF of the form
$1\ts 1+\frac {1}{2} tS +\sum _{i=2}^\infty t^iS_i$.
\endroster
\endproclaim
Thus, equivalence classes of UDF's based on
$\U [[t]]$ are in 1-1 correspondence with unitary
solutions in $\frak g\ts \frak g$
of the classical Yang-Baxter equation. Regarding
(1.17.5), Drinfel'd actually shows more. Namely, he outlines a procedure
which, in principle, constructs a UDF starting from a unitary
solution to the classical Yang-Baxter equation. In practice, however, the
computations necessary
to find the coefficients $F_i$ which comprise the UDF quickly become
insurmountable. In the absence of other techniques to produce UDF's,
it is not surprising that so few are known.

It turns out that there is generally a rich
supply of $\U$-module algebras. First, it is easy to check that
any algebra $A$ which admits an action of $\frak g$ as derivations
is a $\U$-module algebra.
A natural source of this phenomenon comes from group actions. Suppose $G$
is an algebraic which acts
as automorphisms a variety $V$.
Then $\frak g=Lie(G)$ acts as derivations of the coordinate ring
$\Cal O(V)$ of polynomial functions on $V$.
Thus whenever
we have a UDF based on $\U [[t]]$
 the function ring $\Cal O(V)$ will deform.
Now there is a special type of deformation of $\Cal O(V)$ called
a ``star-product'' ($*$-product) with respect to a Poisson bracket,
cf [L]. Recall that a Poisson bracket on $\Cal O(V)$ is a Lie
bracket which satisfies $\{fg,h\}=f\{g,h\}+g\{f,h\}$ for all
$f,g,h\in \Cal O(V)$. A $*$-product is a deformation of $\Cal O(V)$
in which the product $f*g=fg+\sum_{i\geq 0}\mu_i(f,g)$ satisfies
the parity condition $\mu_i(f,g)=(-1)^{i}\mu_i(g,f)$,
the ``null-on-the-constants''
condition $\mu_i(1,f)=0$ for all $i\geq 1$, and is in the ``direction''
of the Poisson bracket in the sense that $\mu_1(f,g)=(1/2)\{f,g\}$.
It is clear that if $F$ is a UDF which satisfies (1.17.3) is used
to deform $\Cal O(V)$ then $r_l(f,g)$ is a Poisson bracket
and the resulting deformation is $*$-product. Hence any quantization
of $\Cal O(V)$ resulting from a UDF is equivalent to a $*$-product
in the direction of some Poisson bracket.
\head{\bf 2. Explicit Formulas}\endhead
The first UDF's we will consider are built from an abelian bialgebra $B$.
In this case, the UDF's
have a remarkably simple form. While widely used, a formal proof of the
following has not appeared in the literature and since it has
extensive applications we include one
for completeness.
\proclaim{Theorem 2.1} Suppose that $k\supset \Bbb Q$. If
$B$ is a commutative bialgebra and $P$ is its
space of primitive elements,
then for any $r\in P\ts P$
$$\exp{(tr)}=\sum_{i=0}^{\infty}{\frac {t^i}{i!}\, r^i}=1\ts 1+t\, r+
\frac {t^2}{2!}\, r^2+\cdots +\frac {t^n}{n!}\, r^n+\cdots$$
is a UDF. \endproclaim
\demo{Proof}
Since $\D_B\ts \I: B\ts B\ra B\ts B\ts B$ is an algebra map, we have
$$[(\D_B\ts \I)(\exp {(tr)})][\exp {(tr)}\ts 1]=
\exp{[ (\D_B \ts \I)(tr)]}\cdot \exp {[tr\ts 1]}\tag 2.2 $$
and
$$[(\I\ts \D_B)(\exp {(tr)})][1\ts \exp {(tr)}]=
\exp{[ (\I \ts \D_B)(tr)]}\cdot \exp {[1\ts tr]}.\tag 2.3$$
Now as $B$ is abelian, (2.2) and (2.3) coincide if and only if
$$(\D_B\ts \I)r +r\ts 1 = (\I \ts \D_B) r +1\ts r$$
which clearly holds whenever $r\in P\ts P$. $\blacksquare$
\enddemo
Up to equivalence, we only need to consider those $r\in P\ts P$
which are skew-symmetric.
In this case, $\exp (tr)$
satisfies (1.17.3) and thus produces $*$-products when used to
deform coordinate rings of algebraic varieties.
\example{Example 2.4}
\roster
\item
The classic use of the exponential deformation formula is to quantize
the coordinate ring $\Cal O(\Bbb R^{2n})=\Bbb R[x_1,\ldots ,x_n,y_1, \ldots
y_n]$
of polynomial functions, (or $C^{\infty}(\Bbb R^{2n})$ in the analytic case),
with respect to the canonical Poisson bracket. For $f,g\in
\Cal O(\Bbb R^{2n})$, this bracket is
$$\{ f,g\}=\frac{1}{2}
\sum _{i=1}^n
\left({\partial {f}\over \partial x_i} {\partial {g}\over \partial y_i}
-{\partial {f}\over \partial y_i}{\partial {g}\over \partial x_i}\right).$$
This Poisson bracket is induced from a translation action
of $\Bbb R^{2n}$ on itself.
Since the derivations ${\partial / \partial x_i}$ and
${\partial / \partial y_i}$ mutually commute,
we can expontiate the Poisson bracket to deform $\Cal O(\Bbb R^{2n})$.
If we
let $f*g$ denote the deformed product,
then it is easy to verify that the quantized
algebra has the Heisenberg relations
$x_i*x_j=x_j*x_i$, $y_i*y_j=y_j*y_i$, and
$x_i*y_j-y_j*x_i=\delta _{ij}$. Note that for any $f,g\in \Cal O(\Bbb R^{2n})$
the product $f*g$ is a polynomial in $t$ and so
it is meaningful to specialize $t$ to any real number.
This construction also
provides a deformation of $k[x_1, \ldots ,x_n,y_1, \ldots y_n]$
over $k[t]$ whenever $k\supset \Bbb Q$.
\item
Consider the commuting derivations
$x_i ({\partial / \partial x_i})$ of $k[x_1,\ldots ,x_n]$. If $k\supset
\Bbb Q$ then
for any scalars $p_{ij}$ we can exponentiate
 $ \sum _{i< j}p_{ij}\,\, x_i ({\partial / \partial x_i})\wedge
x_j ({\partial / \partial x_j})$ to deform
the polynomial ring $k[x_1,\ldots x_n]$. The new relations are
$x_i*x_j=P_{ij}\, x_j*x_i$ where $P_{ij}= \exp (tp_{ij})$. This algebra
is the quantum n-space associated with the standard multi-parameter
quantization of $\Cal O(SL(n))$. If $k=\Bbb R$ or
$\Bbb C$ then for any $f$ and $g$ in $k[x_1,\ldots ,x_n]$, the
deformed product $f*g$ converges for all $t\in k$.\item
Our final application of Theorem 2.2 concerns the enveloping
algebra $U\frak {gl}_2$. Since the matrix units $E_{11}$ and
$E_{22}$ commute, Theorem 1.5 provides a preferred
bialgebra deformation
of $U\frak {gl}_2$ in which
$ {\D '}(a)=
\left[\exp (-t(E_{11}\wedge E_{22})\right](\D(a))
\left[\exp (t(E_{11}\wedge E_{22})\right]$.
It is easy to check that both $E_{11}$ and $E_{22}$ remain
primitive while
$$ {\D'}(E_{12})=E_{12}\ts L^{-1}+L\ts E_{12}\quad
\text {and} \quad  {\D'}(E_{21})=E_{21}\ts L+L^{-1}\ts E_{21}$$
with $L={1\over 2} \exp (t(E_{11}-E_{22}))$.
As this is a preferred deformation, $\D'$ is compatible with the original
multiplication of $\frak {gl}_2$. It is interesting to compare this
with the standard deformation $U_q\frak {gl}_2$ in which the multiplication
must also be changed. \endroster \endexample

In contrast to the commutative case, very little is known about UDF's
based on non-commutative bialgebras. Even for the case of enveloping algebras
of non-abelian Lie algebras,
only one family of UDF's is known.
Most of the remainder of this paper will be
devoted to this family. It is based on an abelian
extension of a Heisenberg Lie algebra. Specifically, let
$k\supset \Bbb Q$ and let $\Cal H\subset
\frak{sl}(n)$ be the $k$-Lie-algebra generated by the diagonal matrices
$H_i =E_{ii}-E_{i+1,i+1}$ for $1\leq i < n $
and elements of the form
$E_{1p}$ or $E_{pn}$ for $p=2,\ldots , n-1$. The element of
$\Cal H\ts \Cal H$ which will serve as the infinitesimal of the UDF is
$$ \frac{1}{2}\left( \sum_{i=1}^{n-1}c_i\, H_i\right)\otimes E_{1n}
\quad + \quad \sum_{i=2}^{n-1}E_{1i}\otimes E_{in} \tag 2.5$$
where $c_1,\ldots c_{n-1}$ are scalars with $c_1+c_{n-1}=1$. The
infinitesimal for the special
case (along with an incorrect UDF) where all $c_i=1/2$ appears both
in [GGS] and [CGS].
Before presenting the correct UDF for the general case,
we need some notation which will be used
for its definition and proof. Let
$$E=E_{1n}, \quad H=\frac{1}{2}\left( \sum_{i=1}^{n-1}c_i\, H_i\right),\quad
{\text {and}}\quad \Cal X =\sum_{i=2}^{n-1}E_{1i}\otimes E_{in}.$$
Also set $c_1=c$ and $c_{n-1}=d$ and so $c+d=1$.
In this notation,  the infinitesimal (2.5)
becomes $H\otimes E+\Cal X $. Now for $m> 0$
define
$$H^{\l m\r }=H(H+1)\cdots (H+m-1)$$
and set $H^{\l 0 \r}=1$.
If $a\in k$ is any scalar, set $H^{\l m\r }_a=(H+a)^{\l m\r }$, i.e.
$$H^{\l m\r }_a=(H+a)(H+a+1)\cdots (H+a+m-1).$$
\proclaim{Theorem 2.6} Let $H_a^{\l m\r}$, $E$, and $\Cal X$ be as
defined above and for each $m\geq 0$ set
$$F_m=\sum_{i=0}^m{m\choose i}\Cal X ^i(H^{\l m-i \r}_{i}\ts E^{m-i}).$$
Then the series
$$F=\sum_{m=0}^{\infty} \frac {t^m}{m!}F_m=1\ts 1+t(H\ts E+\Cal X) +
\frac{t^2}{2!}\left\{ H(H+1)\ts E^2 +2\Cal X (H+1)\ts E +
\Cal X ^2 \right\} +\cdots $$
is a UDF based on $U\Cal H[[t]]$.\endproclaim
For $n=2$ we have $\Cal X =0$ and
$F_1=H\ts E =(1/2) (E_{11}-E_{22})\otimes E_{12}$; in this case,
the corresponding UDF
is the ``quasi-exponential'' formula $\sum _{m=0}^{\infty}(t^m/{m!})
H^{\l m\r}\ts E^m$
of [CGG].
Before proving Theorem
2.6, we first need to establish some elementary formulas.
\proclaim{Lemma 2.7} In the above notation, the following identities hold
for all non-negative integers $r$ and $s$ and all scalars $a$ and $b$:
\roster
\item $H_a^{\l r\r }\, H_{a+r}^{\l s\r }=H_a^{\l r+s\r }.$
\item $H_a^{\l r\r }-H_{a-1}^{\l r\r }=rH_{a}^{\l r-1\r }.$
\item $H_a^{\l r\r }\, E^s=E^s\,H_{( a+s) }^{\l r\r }.$
\item $(E \ts 1)\Cal X =\Cal X (E \ts 1)\qquad {\text {and}}\qquad
(1\ts E)\Cal X =\Cal X (1\ts E).$
\item $\Cal X _{13}\Cal X _{23}-\Cal X _{23}\Cal X _{13}=0
\qquad {\text {and}}\qquad
\Cal X _{13}\Cal X _{12}-\Cal X _{12}\Cal X _{13}=0.$
\item $(1\ts H_a^{\l r\r })\Cal X =\Cal X (1\ts H_{a+d}^{\l r\r })\quad
\text {and}\quad (H_a^{\l r\r }\ts 1)\Cal X =\Cal X (H_{a+c}^{\l r\r }\ts 1).$
\item $\Cal X _{23}\Cal X _{12}^r-\Cal X _{12}^r\Cal X _{23}=
r\, \Cal X _{12}^{r-1}\Cal X _{13}(1\ts E\ts 1).$
\item $(1\ts \D)\Cal X ^r=\sum_{i=0}^r {r\choose i}\Cal X _{12}^i\, \Cal X
_{13}^{r-i}
\qquad {\text {and}}\qquad
(\D \ts 1)\Cal X ^r=\sum_{i=0}^r {r\choose i}\Cal X _{13}^i\, \Cal X
_{23}^{r-i}.$
\item For any scalars $a$ and $b$,
$\D(H^{\l r\r }_a)=\sum_{i=0}^r{r\choose i}H_{b}^{\l  i\r }
\otimes H_{a-b}^{\l r-i\r }.$
\endroster  \endproclaim
\demo{Proof} All of these are straightforward to establish. Properties
(2.7.1) and (2.7.2) are immediate and (2.7.3) through (2.7.6) follow from the
relations
among the generators of the Lie algebra $\Cal H$. Finally, (2.7.7)
through (2.7.9) can each be proved using induction on $r$.
$\blacksquare$\enddemo
\proclaim{Lemma 2.8} If $m\geq 0$ then for all $i\geq 0$
$$\align &\sum_{s=0}^i{i\choose s}\sum_{j=0}^m{m\choose j}
\Cal X _{13}^s\Cal X _{23}^{i-s}
\Cal X _{12}^j(H_{j}^{\l m-j\r }\ts E^{m-j}\ts 1) \\
&=\sum_{s=0}^i{i\choose s}\sum_{j=0}^m{m\choose j}\Cal X _{12}^j\Cal X
_{13}^{s}
\Cal X _{23}^{i-s}(H_{s+j}^{\l  m-j\r }\ts E^{m-j}\ts 1). \tag 2.9
\endalign  $$
\endproclaim
\demo{Proof}
We use induction on $i$. For $i=0$, the assertion is trivial and so
assume it holds for $i$. For $i+1$, the top line of (2.9) becomes,
by (2.7.8),
$$(\Cal X _{13}+\Cal X _{23})\sum_{s=0}^i {i\choose s}\sum_{j=0}^m{m\choose j}
\Cal X _{13}^s\Cal X _{23}^{i-s}
\Cal X _{12}^j(H_{j}^{(m-j\r }\ts E^{m-j}\ts 1).$$
Now, by the inductive hypothesis, this in turn equals
$$(\Cal X _{13}+\Cal X _{23})\sum_{s=0}^i {i\choose s}\sum_{j=0}^m{m\choose j}
\Cal X _{12}^j\Cal X _{13}^{s}
\Cal X _{23}^{i-s}(H_{s+j}^{\l m-j\r }\ts E^{m-j}\ts 1)$$
which, with the aid of (2.7.4), (2.7.5), and (2.7.7) becomes
$$\align
& \sum_{s=0}^i{i\choose s}\sum_{j=0}^m{m\choose j}
\Cal X _{12}^j\Cal X _{13}^{s+1}
\Cal X _{23}^{i-s}(H_{s+j}^{\l m-j\r }\ts E^{m-j}\ts 1)\\
&\quad +\,\, \sum_{s=0}^i{i\choose s}\sum_{j=0}^m{m\choose j}
\Cal X _{12}^j\Cal X _{13}^{s}
\Cal X _{23}^{i+1-s}(H_{s+j}^{\l m-j\r }\ts E^{m-j}\ts 1)\\
&\quad +\,\, \sum_{s=0}^i{i\choose s}\sum_{j=0}^m{m\choose j}
j\Cal X _{12}^{j-1}\Cal X _{13}^{s+1}
\Cal X _{23}^{i-s}(H_{s+j}^{\l m-j\r }\ts E^{m-j+1}\ts 1).\tag 2.10
\endalign $$
For $i+1$, the bottom line of (2.9) is
$$
\sum_{s=0}^{i+1}{{i+1}\choose s}\sum_{j=0}^m{m\choose j}\Cal X _{12}^j\Cal X
_{13}^{s}
\Cal X _{23}^{i+1-s}(H_{s+j}^{\l m-j\r }\ts E^{m-j}\ts 1)$$
which, by the identity ${{i+1}\choose s}={i\choose {s-1}}+{i\choose s}$,
can be expressed as
$$\align
&\sum_{s=0}^{i}{{i}\choose s}
\sum_{j=0}^m{m\choose j}\Cal X _{12}^j\Cal X _{13}^{s+1}
\Cal X _{23}^{i-s}(H_{s+j+1}^{\l m-j\r }\ts E^{m-j}\ts 1)\\
& \quad  +\,\, \sum_{s=0}^{i}{i\choose {s}}\sum_{j=0}^m{m\choose j}\Cal X
_{12}^j\Cal X _{13}^{s}
\Cal X _{23}^{i+1-s}(H_{s+j}^{\l m-j\r }\ts E^{m-j}\ts 1).\tag 2.11
\endalign  $$
Subtracting (2.10) from (2.11) gives and using (2.7.2) gives
$$\align
&\sum_{s=0}^i{i\choose s}\sum_{j=0}^m{m\choose j}
\Cal X _{12}^j\Cal X _{13}^{s+1}
\Cal X _{23}^{i-s}(-(m-j))
(H_{s+j+1}^{\l m-j-1\r })\ts E^{m-j}\ts 1)\\
& \quad +\,\,  \sum_{s=0}^i{i\choose s}\sum_{j=0}^m{m\choose j}
j\Cal X _{12}^{j-1}\Cal X _{13}^{s+1}
\Cal X _{23}^{i-s}(H_{s+j}^{\l m-j\r }\ts E^{m-j+1}\ts 1)\endalign $$
which is the same as
$$\align
&\sum_{s=0}^i{i\choose s}\sum_{j=0}^m{m\choose j}(-(m-j))
\Cal X _{12}^j\Cal X _{13}^{s+1}
\Cal X _{23}^{i-s}
(H_{s+j+1}^{\l m-j-1\r })\ts E^{m-j}\ts 1)\\
&\quad + \,\, \sum_{s=0}^i{i\choose s}\sum_{j=0}^{m-1}{m\choose {j+1}}
(j+1)\Cal X _{12}^{j}\Cal X _{13}^{s+1}
\Cal X _{23}^{i-s}(H_{s+j+1}^{\l m-j-1\r }\ts E^{m-j}\ts 1).
\tag 2.12\endalign  $$
Now (2.12) is equal to zero since ${m\choose j}(m-j)={m\choose {j+1}}(j+1)$.
$\blacksquare$ \enddemo
\demo{Proof of Theorem 2.6}
To verify that $F$ is a UDF, it suffices to show that
$$\sum_{p+q=m}{m\choose p}\left[(1\ts \D)(F_p)\right](1\ts F_q)= \sum_{p+q=m}
{m\choose p}\left[(\D \ts 1)(F_p)\right](F_q\ts 1)\tag 2.13 $$
for all $m$.
The left side of (2.13) is
$$\align
&\left\{\sum_{p+q=m}{m\choose p} \left\{ \sum_{r=0}^p
{p\choose r}(1\ts \D )\Cal X ^ r)(H_{r}^{\l p-r\r }\ts  \D (E^{p-r})\right\}
\right\} \times \\
& \left\{ \sum_{s=0}^q {q\choose s}(1\ts \Cal X ^s)(1\ts H_{s}^{\l q-s\r }
\ts E ^{q-s}) \right\}. \endalign $$
which, using (2.7.3), (2.7.4), and (2.7.8),  can be expressed as
$$\sum_{*} {m\choose p}{p\choose r}{r\choose u}{{p-r}\choose v}{q\choose s}
 \left\{ \Cal X _{12}^u\Cal X _{13}^{r-u}\Cal X _{23}^s
\left( H_{r}^{\l p-r\r }\ts E^v H_{s}^{\l q-s\r } \ts E^{p+q-r-v-s} \right)
\right\} \tag 2.14$$
where $\sum_*$ will indicate that the sum is being taken over
$$p+q=m,\quad 0\leq r\leq p,\quad 0\leq u\leq r,\quad
0\leq v\leq p-r,\quad {\text {and}} \quad 0\leq s \leq q.\tag 2.15$$
In order to compare this with the right side of (2.13),
it will be convenient to re-index its summation by setting
$$u'= s,\quad s'= r-u, \quad p'= q+u+v, \quad
 v'= q-s,\quad r'= u+s, \quad {\text{and}}\quad q'= p-u-v.$$
It is easy to check that the inequalities in (2.15) are equivalent to
having
 $$
p'+q'=m,\quad 0\leq r'\leq p', \quad 0\leq u'\leq r',
\quad
0\leq v'\leq p'-r',\quad {\text{and}} \quad  0\leq s'\leq q'.$$
Hence we may rewrite (2.14) as
$$\sum _{*}{m'\choose p'}{p'\choose r'}{r'\choose
u'}{{p'-r'}\choose v'}{q'\choose s'}
\left\{ \Cal X _{12}^s\Cal X _{13}^{u}\Cal X _{23}^{r-u}
\left( H_{u+s}^{\l q+v-s\r }\ts E^{q-s}
 H_{r-u}^{\l p-v-r\r } \ts E^{p-r} \right)
\right\}.\tag 2.16$$
The right side of (2.13) is
$$\align
&\left\{ \sum_{p+q=m}{m\choose p} \left\{ \sum_{r=0}^p
{p\choose r}((\D\ts 1)\Cal X ^ r)(\D H_{r}^{\l p-r\r }\ts  E^{p-r})\right\}
\right\} \times  \\
& \left\{ \sum_{s=0}^q {q\choose s}( \Cal X ^s \ts 1)(H_{s}^{\l q-s\r }
\ts E^{q-s}\ts 1 )\right\}. \endalign $$
which, by (2.7.3), (2.7.6), (2.7.8), and (2.7.9) becomes
$$\sum _{*} {m\choose p}{p\choose r}{r\choose u}{{p-r}\choose v}{q\choose s}
\Cal X _{13}^u\Cal X _{23}^{r-u}\Cal X _{12}^s
\left( H_{s}^{\l q-s\r }H_{b+sc}^{\l v\r }
\ts E^{q-s}H_{r+sd-b+q-s}^{\l p-r-v\r }
\ts E^{p-r}\right)\tag 2.17$$
where $b$ is any scalar. If we set $b=u+q-sc$ and use the fact that $c+d=1$
together with Lemma 2.8 and (2.7.1), we can rewrite (2.17) as
$$\sum _{*} {m\choose p}{p\choose r} {r\choose u}{{p-r}\choose v}{q\choose s}
\left\{
 \Cal X _{12}^s\Cal X _{13}^{u}\Cal X _{23}^{r-u}
\left( H_{u+s}^{(q+v-s)}\ts E^{q-s}
H_{r-u}^{\l p-r-v\r }\ts E^{p-r} \right)\right\}.\tag 2.18$$
Now (2.16) and (2.18) coincide since
$${m\choose p}{p\choose r}{r\choose u}{{p-r}\choose v}
{q\choose s}={m'\choose p'}{p'\choose r'}{r'\choose
u'}{{p'-r'}\choose v'}{q'\choose s'}$$
and so $F$ is a UDF.
$\blacksquare$ \enddemo

As with the exponential UDF based on commutative bialgebras, the
UDF of Theorem 2.6 can be used to produce deformations of a
large class of algebras and coalgebras.
\example{Example 2.19}
\roster
\item According to remarks following Definition 1.14, the UDF
of Theorem 2.6 provides a preferred bialgebra deformation
of $U\Cal H$ in which $\D'(x)=F^{-1}\D(x)F$ for any $x\in U\Cal H$.
Recall that this comultiplication is compatible with the original
multiplication, unit, and counit of $U\Cal H$. Since we are considering
the Lie algebra $\Cal H$ as a subalgebra of $\frak {gl}(n)$, we can
``extend'' this deformation of $U\Cal H$ to a
(non-standard) deformation of $U\frak {gl}(n)$.
\item Let $\Bbb H\subset GL(n)$ be a simply connected algebraic group
with Lie algebra $\Cal H$. The group $\Bbb H$ acts in a natural way
on $k^n$ and hence $\Cal H$ acts as derivations of
$\Cal O(k^n)=k[x_1,\ldots ,x_n]$ where $E_{ij}$ acts as
$x_i(\partial /\partial  x_j)$. The UDF thus deforms
$k[x_1,\ldots ,x_n]$ and it is not hard to see
that the new relations become
$x_1*x_n-x_n*x_1=(c_1/2)x_1*x_1$ while $x_p*x_n-x_n*x_p=\lambda_p x_1*x_p$
with $\lambda _p =(1/2)(c_p -c_{p-1})+1$ and all other $x_i*x_j-x_j*x_i=0$.
Just like Example 2.4.1, this deformation has the property that
if $f$ and $g$ are in $k[x_1,\ldots ,x_n]$ then the deformed
product $f*g$ is a polynomial in $t$ and so by specialization we obtain
a family of $k$-algebras.
\endroster
\endexample
Even though the UDF of Theorem 2.6 is equivalent to one satisfying
(1.15.3) we do not as of yet have a closed expression
for such a formula. On the positive side, we do have one
for the case $n=2$. In this case, the infinitesimal is
$H\ts E=(1/2)(E_{11}-E_{22}) \ts E_{12}$ and is based on
$U\frak {s}[[t]]$ where $\frak s$, the Lie algebra generated
by $H$ and $E$, is the two-dimensional
solvable Lie algebra, (recall that $[H,E]=E$).
The ``skew-symmetrized''
formula is also based on $U\frak {s}[[t]]$ and has infinitesimal $E\wedge
H=E\ts H-H\ts E$.
\proclaim{Theorem 2.20}
Let $H$ and $E$ be generators of the two-dimensional
solvable Lie algebra where $[H,E]=E$. Then
if
$$F_m=\sum _{r=0}^m (-1)^r {m\choose r}
E^{m-r}H^{\l r\r}\ts E^{r}H^{\l m-r\r } $$
the series
$$F=\sum_{m=0}^{\infty}\frac{t^m}{m!}F_m=1\ts 1+tE\wedge H +
\frac{t^2}{2!}\left( E^2\ts H^{\l 2\r } -2 EH^{\l 1\r }\ts EH^{\l 1\r}
 +H^{\l 2\r }\ts E^2 \right)
+\cdots $$
is a UDF based on $U\frak s[[t]]$.
\endproclaim
\demo{Proof} The result will follow because we can reduce to the case
where
$H$ and $E$ commute and $H^{\l i\r}=H^i$. With these
reductions, $F$ becomes the exponential deformation
formula which, by Theorem 2.1, is a UDF. To obtain the
necessary simplification, we will use the basis $\{E^iH^j\}$ of
$U\frak s$. Note that each $F_m$ is in ``normal form'', that is,
each tensor factor of $F_m$ is expressed using the preferred basis.
To compute
$[(\D\ts 1)(F)](F\ts 1)$ we need must consider
expressions of the form
$$\left[ (\D\ts 1)(E^{p-r}H^{\l r\r }\ts E^{r}H^{\l p-r\r }) \right]
\left(E^{q-s}H^{\l s\r }\ts  E^{s}H^{\l q-s\r }\ts  1\right)$$
or, equivalently,
$$(\D E^{p-r} \ts 1)(\D H^{\l r\r}\ts 1)
(E^{q-s}H^{\l s\r }\ts E^{s}H^{\l q-s\r }\ts 1)
(1\ts 1\ts E^{r}H^{\l p-r\r }).\tag 2.21$$
To write (2.21) in normal form it suffices to know how to do so for
expressions of the form
$$\left( \D H^{\l r\r }\right)
\left(E^{q-s}H^{\l s\r }\ts  E^{s}H^{\l q-s\r }\right)\tag 2.22$$
This can be done using (2.7.9) and (2.7.3) from which we
obtain that, in normal form, (2.22) is
$$\sum_{u=0}^r {r\choose u}E^{q-s}H^{\l s+r-u\r }\ts E^sH^{\l q-s+u\r } .$$
Thus, if $[(\D\ts 1)(F)](F\ts 1)$ is expressed in normal form, the result
is symbolically the same as in the case where $H$ and $E$ commute
and $H^{\l i \r}=H^i$.
Of course, the same is true for
$[(1\ts \D)(F)](1\ts F)$ and so by Theorem 2.1,
$F$ is a UDF based on $U\frak s[[t]]$. $\blacksquare$.
\enddemo
We conclude this paper with some questions about universal deformation
formulas.
\example{Questions 2.23}
\roster
\item Is there a ``conceptual'' way to determine whether a given series
is a UDF? That is, can the direct computational method of proof be avoided?
\item Do the UDF's of Theorem 2.6 and Theorem 2.21 globalize in an
analogous way
that the formula of Example 2.4.1 does? The global version of
this Example 2.4.1 is the DeWilde-Lecomte Theorem which says that if $M$ is
any symplectic manifold then there is a canonical $*$-product
on $C^{\infty}(M)$, see [DL] or, for a geometric proof see [F].
Locally, any such manifold looks
like $\Bbb R^{2n}$ with the symplectic structure given in (2.4.1) and so
the exponential formula gives the local $*$-product. The difficult part
is to show that these local deformations are compatible. Now if $M$ is
just a Poisson manifold and locally a UDF gives a deformation of
$C^{\infty}(M)$, does there exist a global deformation?
\item Are there analogs of UDF's which give strict deformation quantizations
of $C^*$-algebras in the sense of Rieffel (see [R])?
\item Is there a procedure which produces a UDF based on $U\frak g$
from a constant unitary solution to the classical Yang-Baxter
equation based on a Lie algebra $\frak g$? As stated earlier,
there always is such a formula but its explicit form remains a mystery.
\endroster
\endexample

\enddocument